\documentclass[12pt,twoside,english]{article}
\usepackage[T1]{fontenc}
\usepackage[latin1]{inputenc}
\usepackage{geometry}
\usepackage{amsthm}
\usepackage{amsmath}
\usepackage{graphicx}

\makeatletter
\numberwithin{equation}{section}
\numberwithin{figure}{section}

\makeatother

\usepackage{babel}
\begin{document}

\title{The inverse problem in Seismology. Seismic moment and energy of earthquakes.
Seismic hyperbola}

\date{{\normalsize{}Bogdan Felix Apostol }\\
{\normalsize{}Department of Engineering Seismology, Institute of Earth's
Physics, }\\
{\normalsize{}Magurele-Bucharest MG-6, POBox MG-35, Romania }\\
{\normalsize{}email: afelix@theory.nipne.ro}\\
{\normalsize{}{}}}

\maketitle
\relax
\begin{abstract}
The inverse problem in Seismology is tackled in this paper under three
particular circumstances. First, the inverse problem is defined as
the determination of the seismic-moment tensor from the far-field
seismic waves ($P$ and $S$ waves). These waves provide directly
accessible (measurable) experimental data on earthquakes' focal structure
and mechanism. We use the analytical expression of the seismic waves
in a homogeneous isotropic body with a seismic-moment source of tensorial
forces, the source being localized both in space and time. The far-field
waves provide three equations for the sixth unknown parameters of
the general tensor of the seismic moment. Second, the Kostrov vectorial
(dyadic) representation of the seismic moment is used. This representation
relates the seismic moment to the focal displacement in the fault
and the orientation of the fault (moment-displacement relation); it
reduces the seismic moment to four unknown parameters. Third, the
fourth missing equation is derived from the energy conservation and
the covariance condition. In particular, this relation provides access
to the focal volume of the fault and the near-field seismic waves.
The four equations derived here are solved, and the seismic moment
is determined, thus solving the inverse problem in the conditions
described above. It turns out that the seismic moment is traceless,
its magnitude is of the order of the elastic energy stored in the
focal region (as expected), and the solution is governed by the unit
quadratic from associated to the tensor (related to the magnitude
of the longitudinal displacement in the $P$ wave). It is shown that
a useful picture of the seismic moment is the conic represented by
the associated quadratic form, which is a hyperbola (seismic hyperbola).
This hyperbola provides an image for the focal region: its asymptotics
are oriented along the focal displacement and the normal to the fault.
Also, it is shown that the far-field seismic waves allow an estimation
of the volume of the focal region, focal strain, duration of the earthquake
and earthquake energy; the later quantity is a direct measure of the
magnitude of the seismic moment. The special case of an isotropic
seismic moment is presented. 
\end{abstract}
\relax

Running title: \emph{Inverse Seismological Problem} (or \emph{Seismic
Hyperbola})




\section{Introduction}

The inverse problem in Seismology aims at getting information about
the nature and structure of the forces acting in the earthquake's
focus from measurements of the seismic waves at distances far away
from the earthquake focus (at Earth's surface). We present here a
solution to this problem by means of the seismic waves propagating
in a homogeneous isotropic body with localized tensorial forces, the
Kostrov vectorial representation of the seismic moment for a fault
(moment-displacement relation) and the energy conservation together
with the covariance condition. This relation is derived by equating
the energy carried by the far-field seismic waves to the mechanical
work done by forces in the focal region. 

The seismic moment and seismic energy are basic concepts in the theory
of earthquakes.\cite{key-1}-\cite{key-4} The seismic moment has
emerged gradually in the first half of the $20th$ century, the first
estimation of a seismic moment being done by Aki in 1966.\cite{key-5}
The relations between the seismic moment, seismic energy, the mean
displacement in the focal region, the rate of the seismic slip and
the earthquake magnitude are recognized today as very convenient tools
for characterizing the earthquakes.\cite{key-6}-\cite{key-8} 

The inverse (inversion) problem\cite{key-9} is solved usually by
determining the seismic-moment components $M_{ij}$ ($i,\, j=1,2,3$)
from information provided by far-field seismic waves at different
locations and times,\cite{key-10}-\cite{key-14} or free oscillations
of the earth, long-period surface waves, supplemented, in general,
with additional relevant information (constraints; see Ref. \cite{key-15}
and references therein). Besides noise, the information provided by
such data may reflect particularities of the structure of the focal
region and the focal mechanism which are not included, usually, in
equations, like the structure factor of the focal region, both spatial
and temporal, or deviations from homogeneity and isotropy. In particular,
waves measured at different locations (or times) may lead to overdetermined
systems of equations for the unknowns $M_{ij},$ and the solutions
must be \textquotedbl{}compatibilized\textquotedbl{}. A proper procedure
of compatibilization may lead, in fact, to redundant equations, if
the covariance of the equations is not ensured. Indeed, the experimental
data may often be used in a non-covariant form, which makes the results
dependent on the reference frame. The covariance is understood in
this paper as the invariance of the form of the equations to translations
and rotations (independence of the reference frame). We may add that
the normal modes of the pure free oscillations do not imply a source
of waves, while surface waves, having sources on the surface, have
a very indirect connection to the body waves generated in the focal
region. Surface displacement in the main shock of an earthquake is
often used, which has a very indirect relevance for the earthquake
source and mechanism. 

We present here a direct way of determining (analytically) the seismic
moment for a shear faulting (as well as for an isotropic source) by
using the far-field waves generated by a time-localized tensorial
point source. The waves produced by extended sources imply additional
information regarding the spatial and temporal structure factors;
the inverse problem in this case is a more complex problem, which
remains beyond the aim of the present paper.

We consider that the available data are the displacement vectors produced
by the seismic waves in the wave region. The information provided
by these data is the magnitude of the longitudinal ($P$-wave) displacement
(one parameter) and the transverse-wave displacement vector ($S$-wave,
two parameters; we assume that the direction of the earthquake focus
is known). These data provide three independent parameters, related
to the components of the seismic moment by three equations. They may
be viewed as a minimal set of independent data. It follows that, restricting
ourselves to these data only, the seismic moment has only three independent
components. On the other hand, according to Kostrov representation,
the seismic moment is characterized by its magnitude and the fault
orientation and the fault slip, which are two mutualy perpendicular
unit vectors. This information includes four independent parameters.
We can see, on one hand, according to Kostrov representation, that
only four out of six components of the seismic moment are independent
and, on the other hand, we need a fourth equation in order to determine
the four independent components of the seismic focus. We provide in
this paper the fourth equation, which is the equation of energy conservation
together with the covariance condition. The covariance condition reduces
the four independent components of the seismic moment to three, which
makes possible the determination of the seismic moment from the seismic-wave
displacement. Also, we show that an image of the forces acting in
the focal region and the geometry of the fault can be obtained by
a so-called \textquotedbl{}seismic hyperbola\textquotedbl{}. 

It is widely assumed that typical tectonic earthquakes originate in
a localized focal region, with dimensions much shorter than the distance
to the observation point (and the seismic wavelengths). The tensorial
seismic force density
\begin{eqnarray}
F_{i} & = & M_{ij}\partial_{j}\delta(\mathbf{R}-\mathbf{R}_{0})\label{1}
\end{eqnarray}
 is used for the seismic focus,\cite{key-2,key-4,key-16} where $M_{ij}$
is the tensor of the seismic moment, $\delta$ is the Dirac delta
function and $\mathbf{R}_{0}$ is the position of the focus (hypocentre).
We assume that the position $\mathbf{R}_{0}$ is a known parameter.
The labels $i,\, j$ denote the Cartesian axes and summation over
repeating suffixes is assumed (throughout this paper). The seismic
tensor $M_{ij}$ is a symmetric tensor, which, in general, has six
independent components. It may be decomposed into double-couple (shear
faulting) and dipole components and an isotropic component; departure
from double-couple components reflects a complex shear faulting, tensile
faulting, volcanic morphology, etc.\cite{key-15},\cite{key-17}-\cite{key-21}
The force given by equation (\ref{1}) is a generalization of the
double-couple representation of the seismic force. Indeed, let us
assume a force density $\mathbf{F}(\mathbf{R})=\mathbf{f}g(\mathbf{R})$,
where $\mathbf{f}$ is the force and $g(\mathbf{R})$ is a distribution
function; a point couple associated with a force acting along the
$i$-th direction can be represented as
\begin{equation}
\begin{array}{c}
f_{i}g(x_{1}+h_{1},x_{2}+h_{2},x_{3}+h_{3})-f_{i}g(x_{1},x_{2},x_{3})\simeq\\
\\
\simeq f_{i}h_{j}\partial_{j}g(x_{1},x_{2},x_{3})\:\:\:,
\end{array}\label{2}
\end{equation}
 where $h_{j}$, $j=1,\,2,\,3$, are the components of an infinitesimal
displacement $\mathbf{h}$; $x_{i}$, $i=1,\,2,\,3$, are the coordinates
of the position $\mathbf{R}$ and $\partial_{j}$ denotes the derivative
with respect to $x_{j}$. The force moment (torque) $t_{ij}=f_{i}h_{j}$
is generalized in equation (\ref{2}) to a symmetric tensor $M_{ij}$,
which is the seismic moment entering equation (\ref{1}); in addition,
the distribution $g(\mathbf{R})$ can be replaced by $\delta(\mathbf{R}-\mathbf{R}_{0})$
for a spatially localized focal region. The $\delta$-function used
in equation (\ref{1}) is an approximation for the shape of the focal
region. In equation (\ref{1}) the focus is viewed as being localized
over a distance of order $l$ (volume of order $l^{3}$), much shorter
than the distance $R$ to the observation point ($l\ll R$). 

The seismic moment depends on the time $t$; we may write $M_{ij}(t)=M_{ij}h(t)$,
where $h(t)$ is a positive function, localized at $t=0$, which includes
the time dependence of the seismic moment; we assume $max[h(t)]=h(0)=1$
and denote by $T$ the (short) duration of the seismic event; the
time $T$ is much shorter than any time of interest, such that we
may view the function $h(t)$ as being represented by $T\delta(t)$.
The particular case $h(t)=T\delta(t)$ is called an elementary earthquake
in Refs. \cite{key-16}. (The function $h(t)$ should not be mistaken
for the magnitude of the displacement vector $\mathbf{h}$ used above). 

For a homogeneous isotropic body the seismic waves generated by the
tensorial force given by equation (\ref{1}) are governed by the equation
of the elastic waves
\begin{eqnarray}
\ddot{u}_{i}-c_{t}^{2}\Delta u_{i}-(c_{l}^{2}-c_{t}^{2})\partial_{i}div\mathbf{u} & = & \frac{1}{\rho}M_{ij}(t)\partial_{j}\delta(\mathbf{R})\:\:\:,\label{3}
\end{eqnarray}
where $u_{i}$ are the components of the displacement vector $\mathbf{u}$,
$c_{l,t}$ are the velocities of the longitudinal and tranverse waves,
respectively, $\rho$ is the density and $\mathbf{R}$ is the position
vector drawn from the focus (taken as the origin of the reference
frame) to the observation point. The solution of this equation\cite{key-2,key-4,key-16}
can be written as $\mathbf{u}=\mathbf{u}^{n}+\mathbf{u}^{f}$, where
\begin{equation}
\begin{array}{c}
u_{i}^{n}=-\frac{1}{4\pi\rho c_{t}^{2}}\frac{M_{ij}x_{j}}{R^{3}}h(t-R/c_{t})+\\
\\
+\frac{1}{8\pi\rho R^{3}}\left(M_{jj}x_{i}+4M_{ij}x_{j}-\frac{9M_{jk}x_{i}x_{j}x_{k}}{R^{2}}\right)\cdot\\
\\
\cdot\left[\frac{1}{c_{l}^{2}}h(t-R/c_{l})-\frac{1}{c_{t}^{2}}h(t-R/c_{t})\right]
\end{array}\label{4}
\end{equation}
 is the near-field displacement ($R$ comparable with $l$) and 
\begin{equation}
\begin{array}{c}
u_{i}^{f}=-\frac{1}{4\pi\rho c_{t}^{3}}\frac{M_{ij}x_{j}}{R^{2}}h^{'}(t-R/c_{t})-\frac{1}{4\pi\rho}\frac{M_{jk}x_{i}x_{j}x_{k}}{R^{4}}\cdot\\
\\
\cdot\left[\frac{1}{c_{l}^{3}}h^{'}(t-R/c_{l})-\frac{1}{c_{t}^{3}}h^{'}(t-R/c_{t})\right]
\end{array}\label{5}
\end{equation}
 is the far-field displacement ($R\gg l$). The near-field region
is defined by distances $R$ of the order $l$, while the far-field
region is defined by distances $R$ much larger than $l$. The short
duration $T$ of the seismic event (duration of activity of the focus)
enters equations (\ref{4}) and (\ref{5}) through $h(t)$ and the
derivative $h^{'}(t)$, which is of the order $1/T$. The displacement
vectors given by equations (\ref{4}) and (\ref{5}) include the longitudinal
wave (denoted by suffix $l$, not to be confused with length $l$),
propagating with velocity $c_{l}$, and the transverse wave (suffix
$t$), propagating with velocity $c_{t}$; in the far-field region
the displacement vectors of the longitudinal wave ($P$ wave) and
the transverse wave ($S$ wave) are mutually orthogonal (this is not
so for the $l,\, t$-waves in the near-field region). As long as the
function $h(t)$ may be viewed as a localized function, the magnitude
of the displacement vectors varies as $1/R^{2}$ for the near-field
wave and $1/R$ for the far-field waves. Their direction is determined
by the tensor of the seismic moment $M_{ij}$ (in particular the vector
with components $M_{ij}x_{j}$). The far-field waves given in equation
(\ref{5}) are shell spherical waves with a thickness of the order
$\Delta R\simeq c_{l,t}T$. A superposition of forces given by equation
(\ref{1}), localized at different positions $\mathbf{R}_{0}$ and
different times, corresponds to a structured focus, and the elementary
displacement given by equations (\ref{4}) and (\ref{5}) gives access
to the structure factor of the focal region.\cite{key-16}

\section{Far-field seismic waves}

It is convenient to introduce the notations 
\begin{equation}
M_{i}=M_{ij}n_{j}\:\:,\:\: M_{0}=M_{ii}\:\:,\:\: M_{4}=M_{ij}n_{i}n_{j}\:\:\:,\label{6}
\end{equation}
where $\mathbf{n}$ is the unit vector along the radius drawn from
the focus to the observation point (observation radius), $x_{i}=Rn_{i}$,
and $h_{l,t}=h(t-R/c_{l,t})$; henceforth we consider the unit vector
$\mathbf{n}$ a known vector. $M_{0}$ is the trace of the seismic-moment
tensor and $M_{4}$ is the quadratic form associated to the seismic-moment
tensor, constructed with the unit vector $\mathbf{n}$; we call it
the unit quadratic form of the tensor. The vector $\mathbf{M}$ can
be called the \textquotedbl{}projection\textquotedbl{} of the tensor
along the focus-observation point direction (observation direction). 

Making use of these notations, the seismic waves given by equations
(\ref{4}) and (\ref{5}) can be decomposed into $l$- and $t$-waves,
written as $\mathbf{u}^{n}=\mathbf{u}_{l}^{n}+\mathbf{u}_{t}^{n}$,
\begin{equation}
\begin{array}{c}
\mathbf{u}_{l}^{n}=\frac{h_{l}}{8\pi\rho c_{l}^{2}R^{2}}\left[(M_{0}-9M_{4})\mathbf{n}+4\mathbf{M}\right]\:\:,\\
\\
\mathbf{u}_{t}^{n}=-\frac{h_{t}}{8\pi\rho c_{t}^{2}R^{2}}\left[(M_{0}-9M_{4})\mathbf{n}+6\mathbf{M}\right]\:\:\:,
\end{array}\label{7}
\end{equation}
 and $\mathbf{u}^{f}=\mathbf{u}_{l}^{f}+\mathbf{u}_{t}^{f}$, 
\begin{equation}
\mathbf{u}_{l}^{f}=-\frac{h_{l}^{'}}{4\pi\rho c_{l}^{3}R}M_{4}\mathbf{n}\:\:,\:\:\mathbf{u}_{t}^{f}=\frac{h_{t}^{'}}{4\pi\rho c_{t}^{3}R}\left(M_{4}\mathbf{n}-\mathbf{M}\right)\:\:.\label{8}
\end{equation}
 For numerical purposes we take the \textquotedbl{}maximum deviation\textquotedbl{}
of the near-field diplacement $\mathbf{u}_{l,t}^{n}$ (with its sign)
for $t=R/c_{l,t}$, \emph{i.e. }we take $h_{l,t}(0)=1$. Equally well,
we can take the average values of the vectors $\mathbf{u}_{l,t}^{n}$
over the support $T$ of the functions $h_{l,t}$, or $\Delta R$,
which is of the order $c_{l,t}T$. Henceforth, $h_{l,t}$ in equations
(\ref{7}) are understood as $h_{l,t}(0)=1$. The functions $h_{l,t}^{'}$
are scissor-like functions (\textquotedbl{}double-shock\textquotedbl{}
functions), with two sides with opposite signs (corresponding to $t>0$
or $t<0$), extending over $T$, or the distance $\Delta R$; their
\textquotedbl{}maximum deviations\textquotedbl{} are of the order
$\pm1/T$; for numerical estimations it is convenient to introduce
the notations $\mathbf{v}_{l,t}=\mathbf{u}_{l,t}^{f}/Th_{l,t}^{'}$
and take the \textquotedbl{}maximum deviation\textquotedbl{} of these
functions (with their sign), on any side of the functions $h_{l,t}^{'}$,
the same side for $\mathbf{v}_{l}$ and $\mathbf{v}_{t}$ ($\mathbf{v}_{l,t}$
may depend on the side of the functions $h_{l,t}^{'}$, since the
functions $h_{l,t}(t)$ are not necessarily symmetric with respect
to $t=0$). Similarly, we can take the average values of $\mathbf{v}_{l,t}$
over any side of the functions $h_{l,t}^{'}$ (the same for $\mathbf{v_{l}}$
and $\mathbf{v}_{t})$. The displacement vectors $\mathbf{v}_{l,t}$
are directly accessible experimentally. We consider them as data for
our problem. Making use of these notations, equations (\ref{8}) become
\begin{equation}
\mathbf{v}_{l}=-\frac{1}{4\pi\rho Tc_{l}^{3}R}M_{4}\mathbf{n}\:\:,\:\:\mathbf{v}_{t}=\frac{1}{4\pi\rho Tc_{t}^{3}R}\left(M_{4}\mathbf{n}-\mathbf{M}\right)\:\:.\label{9}
\end{equation}
 We note that the vectors $R^{2}\mathbf{u}_{l,t}^{n}$ and $R\mathbf{v}_{l,t}$
depend on the density $\rho$, the duration $T$, the seismic moment
and the elastic coefficients of the body (velocities of the elastic
waves); if local deviations from this pattern are observed, the body
is not locally homogeneous and isotropic (or the focus is not localized). 

The displacement in the far-field waves is determined by three independent
parameters: the magnitude of the vectors $\mathbf{v}_{l,t}$ (two
parameters) and the direction of the transverse vector $\mathbf{v}_{t}$
(one parameter). Consequently, we may view the equations 
\begin{equation}
\mathbf{M}=-4\pi\rho TR\left(c_{l}^{3}\mathbf{v}_{l}+c_{t}^{3}\mathbf{v}_{t}\right)\:\:\:,\label{10}
\end{equation}
 derived from equations (\ref{8}), as three independent equations
for the six unknown components $M_{ij}$ of the seismic moment; by
multipling by $n_{i}$ and summing over $i$, we get the first equation
(\ref{8}),
\begin{equation}
M_{4}=M_{ij}n_{i}n_{j}=-4\pi\rho TRc_{l}^{3}(\mathbf{v}_{l}\mathbf{n})=-4\pi\rho TRc_{l}^{3}v_{l}\:\:\:,\label{11}
\end{equation}
which is not independent of the three equations written above. We
view $\mathbf{v}_{l,t}$ as (known) quantities measured experimentally,
and $\rho$, $R$, $c_{l,t}$ as known parameters; duration $T$ will
be determined shortly. A simple observation would show that for given
displacements $\mathbf{v}_{l,t}$ and given $T$ we may solve equations
(\ref{10}) and get the three independent components of the seismic
moment $M_{ij.}$. Unfortunately, leaving aside that the other three
components are left as free parameters by such a procedure, the measurement
of the duration $T$ from $\Delta r/c_{l,t}$, where $\Delta r$ is
the projection of $\Delta R$ on Earth's surface, is dependent on
the local frame, and, consequently, would not provide a suitable input
data for covariant equations. 

We note in equations (\ref{10}) and (\ref{11}) the consistency (compatibility)
relation $M^{2}>M_{4}^{2}$, derived from $v_{t}^{2}>0$ ($v_{l,t}$
denote the magnitudes of the vectors $\mathbf{v}_{l,t}$). The inverse
problem discussed in this paper consists in determining the tensor
$M_{ij}$ from the displacement $\mathbf{v}_{l,t}$ in the far-field
waves, making use of additional, model-related, information. The model
we use is provided by the fault geometry of the focal zone. We can
see that only three components of the seismic moment $M_{ij}$ are
independent. We determine the seismic-moment tensor by means of the
vectors $\mathbf{M}$ and $\mathbf{n}$ (experimentally accessible).
The special case of an isotropic moment is presented. We note that
equations (\ref{9}) are manifestly covariant. Also, we note that
having known $\mathbf{M}$ and $M_{4}$ we can have access to the
near-field diplacement given by equations (\ref{7}), provided we
know $M_{0}$.

\section{Energy of earthquakes}

If we multiply equation (\ref{3}) by $\dot{u}_{i}$ and sum over
the suffix $i$, we get the law of energy conservation
\begin{equation}
\begin{array}{c}
\frac{\partial}{\partial t}\left[\frac{1}{2}\rho\dot{u}_{i}^{2}+\frac{1}{2}\rho c_{t}^{2}(\partial_{j}u_{i})^{2}+\frac{1}{2}\rho(c_{l}^{2}-c_{t}^{2})(\partial_{i}u_{i})^{2}\right]-\\
\\
-\rho c_{t}^{2}\partial_{j}(\dot{u}_{i}\partial_{j}u_{i})-\rho(c_{l}^{2}-c_{t}^{2})\partial_{j}(\dot{u}_{j}\partial_{i}u_{i})=\dot{u}_{i}M_{ij}(t)\partial_{j}\delta(\mathbf{R})\:\:.
\end{array}\label{12}
\end{equation}
According to this equation, the external force performs a mechanical
work in the focus ($\dot{u}_{i}M_{ij}(t)\partial_{j}\delta(\mathbf{R})$
per unit volume and unit time). The corresponding energy is transferred
to the waves (the term in the square brackets in equation (\ref{12})),
which carry it through the space (the term including the $div$ in
equation (\ref{12})). It is worth noting that outside the focal region
the force is vanishing. Also, the waves do not exist inside the focal
region. Therefore, limiting ourselves to the displacement vector of
the waves, we have not access to the mechanical work done by the external
force in the focal region. This circumstance arises from the localized
character of the focus.

In the far-field region we can use the decomposition $\mathbf{u}=\mathbf{u}_{l}+\mathbf{u}_{t}$
in longitudinal and transverse waves, where $curl\mathbf{u}_{l}=0$
and $div\mathbf{u}_{t}=0$; this decomposition leads to
\begin{equation}
\frac{\partial e_{l,t}}{\partial t}+c_{l,t}div\mathbf{s}_{l,t}=0\:\:\:,\label{13}
\end{equation}
 where
\begin{equation}
e_{l,t}=\frac{1}{2}\rho\left(\dot{\mathbf{u}}_{l,t}^{f}\right)^{2}+\frac{1}{2}\rho c_{l,t}^{2}\left(\partial_{i}u_{l,tj}^{f}\right)^{2}\label{14}
\end{equation}
` is the energy density and 
\begin{equation}
s_{l,ti}=-\rho c_{l,t}\dot{u}_{l,tj}^{f}\partial_{i}u_{l,tj}^{f}\label{15}
\end{equation}
 are the components of the energy flux densities per unit time (the
flow vectors). From equation (\ref{13}) we can see that the energy
is transported with velocities $c_{l,t}$ (as it is well known). The
volume energy $E=\int d\mathbf{R}(e_{l}+e_{t})$ is equal to the total
energy flux
\begin{equation}
\Phi=-\int dtd\mathbf{R}\left(c_{l}div\mathbf{s}_{l}+c_{t}div\mathbf{s}_{t}\right)=-\int dt\oint d\mathbf{S}\left(c_{l}\mathbf{s}_{l}+c_{t}\mathbf{s}_{t}\right)\:\:.\label{16}
\end{equation}
 Making use of equations (\ref{8}) and taking $h^{''}=-1/T^{2}$
as an order-of-magnitude estimate, we get 
\begin{equation}
E=\Phi=\frac{4\pi\rho}{T}R^{2}\left(c_{l}v_{l}^{2}+c_{t}v_{t}^{2}\right)\:\:;\label{17}
\end{equation}
this relation gives the energy released by the earthquake in terms
of the displacement measured in the far-field region and the (short)
duration of the earthquake. From equations (\ref{9}) we get the relation
\begin{equation}
E=\frac{1}{4\pi\rho c_{t}^{5}T^{3}}\left[M^{2}-\left(1-c_{t}^{5}/c_{l}^{5}\right)M_{4}^{2}\right]\label{18}
\end{equation}
between energy and the seismic moment.

\section{Geometry of the focal region }

\begin{figure}
\noindent \begin{centering}
\includegraphics[clip,scale=0.8]{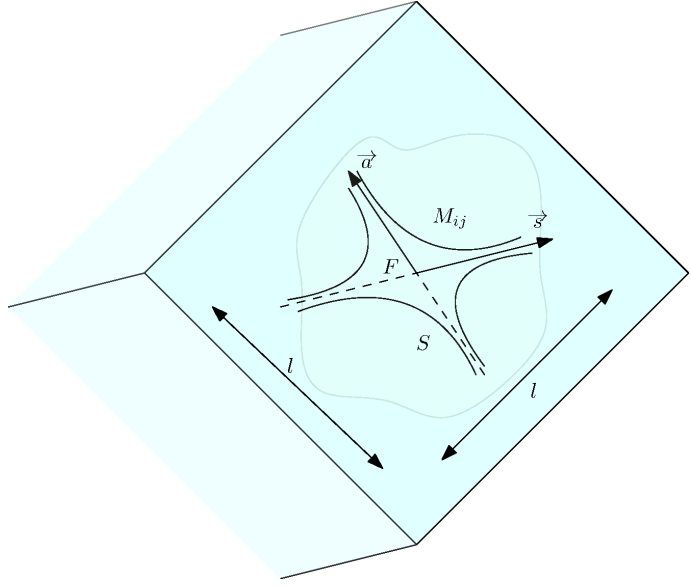}
\par\end{centering}

\caption{A fault focal cross-section with area $S$ (dimension $l$, focus
$F$); $\mathbf{s}$ is the unit vector normal to the fault and $\mathbf{a}$
is the unit vector of the focal displacement (in the plane of the
fault); the seismic-moment tensor $M_{ij}$ is represented by the
rectangular hyperbola with the axes along the vectors $\mathbf{s}$
and $\mathbf{a}$. \label{fig:Fig.1} }

\end{figure}

Let us consider a point torque $t_{ij}=f_{i}h_{j}$, where $h_{j}$
are viewed as infinitesimal distances and $f_{i}$ denote the components
of a force $\mathbf{f}$; the force $\mathbf{f}$ originates in a
volume force density $\partial_{j}\sigma_{ij}$, where $\sigma_{ij}$
is the stress tensor; the latter can be expressed as $\sigma_{ij}=2\mu u_{ij}+\lambda u_{kk}\delta_{ij}$,
where $\mu$ and $\lambda$ are the Lame coefficients ($c_{l}^{2}=(2\mu+\lambda)/\rho$,
$c_{t}^{2}=\mu/\rho$), $u_{ij}=\frac{1}{2}(\partial_{j}u_{i}$+$\partial_{i}u_{j})$
is the strain tensor and $\mathbf{u}$, with components $u_{i}$,
is the displacement vector.\cite{key-22} We can write
\begin{equation}
\begin{array}{c}
t_{ij}=f_{i}h_{j}=\int d\mathbf{r}\partial_{k}\sigma_{ik}\cdot h_{j}=\\
\\
=\mu\int d\mathbf{r}\partial_{k}^{2}u_{i}\cdot h_{j}+(\mu+\lambda)\int d\mathbf{r}\partial_{k}\partial_{i}u_{k}\cdot h_{j}=\\
\\
=\mu\oint dS\cdot s_{k}\partial_{k}u_{i}\cdot h_{j}+(\mu+\lambda)\oint dS\cdot s_{k}\partial_{i}u_{k}\cdot h_{j}\:\:\:,
\end{array}\label{19}
\end{equation}
where the $\mathbf{r}$-integration is performed over the focal volume
surrounded by the surface $S$ and $\mathbf{s}$ is the unit vector
normal to this surface. We may write $\partial_{i}u_{k}\simeq\Delta u_{k}/\Delta x_{i}$
for the derivatives of $u_{k}$ and use $\frac{\Delta u_{k}}{\Delta x_{i}}\cdot h_{j}=\Delta u_{k}\delta_{ij}=u_{k}\delta_{ij}$,
where $u_{k}$ is the displacement on the surface. These equalities
follow from the point-like nature of the torque. We note that $\mathbf{u}$
here is the focal displacement, which is distinct from the displacement
in the waves. It follows
\begin{equation}
t_{ij}=\mu S\cdot\overline{s_{j}u_{i}}+(\mu+\lambda)S\cdot\overline{s_{k}u_{k}}\delta_{ij}\:\:\:,\label{20}
\end{equation}
where the overbar denotes the average over the surface with area $S$.
This relation acquires a useful form for a localized (plane) fault.
We assume that the fault focal region includes two plane-parallel
surfaces, each with (small) area $S$, separated by a (small) distance
$d$, sliding against one another. The focal area is determined by
two lengths $l_{1,2}$, $S=l_{1}l_{2}$. In general, the lengths $l_{1},\, l_{2},\, d$
are distinct; in order to ensure the compatibility with the localization
provided by the $\delta$-function (used in deriving the waves), we
assume $l_{1}=l_{2}=d=l$. For such a model of localized fault the
product $\overline{s_{j}u_{i}}$ may be replaced by $2s_{j}\overline{u}_{i}$,
where the vector $\mathbf{s}$ is the unit vector normal to the fault
(we note that the integration over the surfaces perpendicular to the
fault is zero, due to the opposing (sliding) displacements). In view
of the small extension of the focal region, we may drop the average
bar over $u_{i}$. In addition, this model of fault-slip implies $s_{k}u_{k}=0$,
\emph{i.e.} the normal to the fault $\mathbf{s}$ and the focal displacement
(fault slip) $\mathbf{u}$ are mutually orthogonal vectors. In order
to distinguish the focal displacement from the displacement in the
seismic waves, we attach the superscript $0$ to the focal displacement.
The seismic moment is obtained by symmetrizing the expression given
by equation (\ref{20}); we get
\begin{equation}
M_{ij}=2\mu S\left(s_{i}u_{j}^{0}+s_{j}u_{i}^{0}\right)=2\mu Su^{0}\left(s_{i}a_{j}+a_{i}s_{j}\right)\:\:\:,\label{21}
\end{equation}
where we introduce the unit vector $\mathbf{a}$ along the direction
of the focal displacement; we write $u_{i}=u^{0}a_{i}$, where $u^{0}$
is the magnitude of the focal displacement and $a_{i}^{2}=1$. We
can see that the seismic moment is represented in equation (\ref{21})
by two orthogonal vectors ($\mathbf{a}\mathbf{s}=0$): the unit vector
$\mathbf{a}$ along the focal displacement $\mathbf{u}^{0}$ and the
unit vector $\mathbf{s}$, which gives the orientation of the fault.
This is the moment-displacement relation derived by Kostrov\cite{key-7,key-8}
for the slip along a (point-like) fault surface (see also Refs. \cite{key-2,key-4});
it can be called a vectorial, or dyadic, representation of the seismic
moment. We note the invariant $M_{0}=M_{ii}=0$, which tells that
the seismic moment in this representation is a traceless tensor. This
particularity gives access to the near-field waves (equations (\ref{7})),
which become
\begin{equation}
\mathbf{u}_{l}^{n}=\frac{h_{l}}{8\pi\rho c_{l}^{2}R^{2}}\left(4\mathbf{M}-9M_{4}\mathbf{n}\right)\:\:,\:\:\mathbf{u}_{t}^{n}=-\frac{3h_{t}}{8\pi\rho c_{t}^{2}R^{2}}\left(2\mathbf{M}-3M_{4}\mathbf{n}\right)\label{22}
\end{equation}
 ($\mathbf{M}$ and $M_{4}$ are given by equations (\ref{10}) and
(\ref{11})). In addition, we note the relations $M_{4}^{0}=M_{ij}s_{i}s_{j}=0$
and $M_{i}^{0}=M_{ij}s_{j}=2\mu Su^{0}a_{i}$; the former relation
shows that the quadratic form associated to the seismic moment in
the focal region is degenerate (it is represented by a conic), while
the latter relation shows that the \textquotedbl{}force\textquotedbl{}
in the focal region is directed along the focal displacement; both
relations are expected from the Kostrov construction of the tensor
of the fault seismic moment (Fig. \ref{fig:Fig.1}). 
 The relations $M_{0}=0$ and $M_{4}^{0}=0$ reduce the number of
independent parameters of the tensor $M_{ij}$ from six to four. 

\begin{figure}
\noindent \begin{centering}
\includegraphics[clip,scale=0.6]{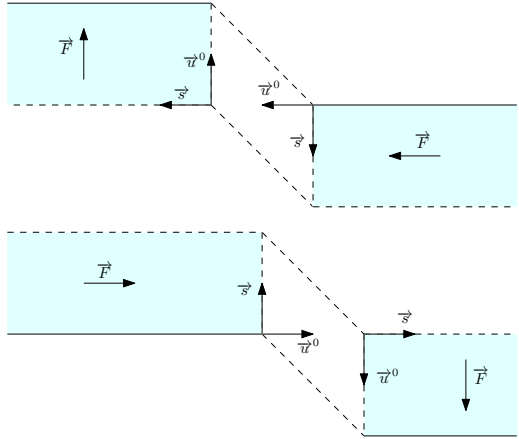}
\par\end{centering}

\caption{Two couples of sliding displacements ($\mathbf{u}^{0}$) and two orthogonal
orientations ($\mathbf{s}$) in a fault focal region, illustrating
the indeterminacy in the Kostrov construction of the seismic moment;
$F$ denote the forces which give the torque. \label{fig:Fig. 2}}

\end{figure}

It is worth noting an uncertainty (indeterminacy) of the dyadic construction
of the seismic-moment tensor. We can see from equation (\ref{21})
that the seismic moment is invariant under the inter-change $\mathbf{s}\longleftrightarrow\mathbf{a}$.
This means that from the knowledge of the seismic moment $M_{ij}$
we cannot distinguish between the two orthogonal vectors $\mathbf{s}$
and $\mathbf{a}$ (fault direction and fault slip). Another symmetry
of the seismic moment given by equation (\ref{21}) is $\mathbf{s}\longleftrightarrow-\mathbf{a}$
(and $\mathbf{s}\longleftrightarrow-\mathbf{s}$, $\mathbf{a}\longleftrightarrow-\mathbf{a}$),
which means that we cannot distinguish between the signs of the vectors
$\mathbf{s}$ and $\mathbf{a}$ (as expected from the construction
of the seismic moment in equation (\ref{21})); this uncertainty is
shown in Fig. \ref{fig:Fig. 2}. 

In equation (\ref{21}) the seismic moment is determined by four parameters:
three components of the displacement vector $\mathbf{u}^{0}$ and
one component of the (transverse) unit vector $\mathbf{s}$. By using
this vectorial representation, the number of independent parameters
of the seismic moment is reduced from six to four. We have, up to
this moment, only the three equations (\ref{10}) for these unknown
parameters. The considerations made above for the vectorial representation
of the seismic moment provides a fourth equation, relating the mechanical
work $W$ done in the focal region to the magnitude of the focal diplacement.

Indeed, from equation (\ref{12}) the mechanical work in the focal
region is given by
\begin{equation}
W=\int dt\int d\mathbf{R}\dot{u}_{i}^{0}(t)M_{ij}(t)\partial_{j}\delta(\mathbf{R})\:\:;\label{23}
\end{equation}
we may assume $\dot{u}_{i}^{0}(t)=\dot{h}(t)u_{i}^{0}$, and, since
$M_{ij}(t)=M_{ij}h(t)$, we get 
\begin{equation}
W=\frac{1}{2}\int d\mathbf{R}u_{i}^{0}M_{ij}\partial_{j}\delta(\mathbf{R})\:\:.\label{24}
\end{equation}
 In this equation we may view the function $\delta(\mathbf{R})$ as
corresponding to the shape of the focal surface, such that we may
replace $\partial_{j}\delta(\mathbf{R})$ by $s_{j}/l^{4}$; using
$V=l^{3}$ for the focal volume, we get $W\simeq\frac{1}{2l}u_{i}^{0}M_{ij}s_{j}$.
Here, we may take approximately $u^{0}$ for $l$, which leads to
$W\simeq\frac{1}{2}a_{i}M_{ij}s_{j}$. Therefore, making use of equation
(\ref{21}), we get $W\simeq\mu Su^{0}=\mu V$; we can see that the
mechanical work done in the focal region is of the order of the elastic
energy stored in the focal region, as expected. By equating $W$ with
energy $E$ (and $\Phi$) given by equation (\ref{17}), the fourth
equation
\begin{equation}
\mu V=\frac{4\pi\rho}{T}R^{2}\left(c_{l}v_{l}^{2}+c_{t}v_{t}^{2}\right)\label{25}
\end{equation}
 is obtained; it can also be written as
\begin{equation}
V=\frac{4\pi}{c_{t}^{2}T}R^{2}\left(c_{l}v_{l}^{2}+c_{t}v_{t}^{2}\right)\:\:.\label{26}
\end{equation}
This equation gives the volume of the focal region in terms of the
displacement in the far-field seismic waves (provided duration $T$
is known); the seismic moment given by equation (\ref{21}) can be
written as
\begin{equation}
M_{ij}=2\mu V\left(s_{i}a_{j}+a_{i}s_{j}\right)\:\:\:,\label{27}
\end{equation}
where $V$ can be inserted from equation (\ref{26}). It remains to
determine the vectors $\mathbf{a}$ and $\mathbf{s}$ by using equations
(\ref{10}) and the covariance condition, in order to solve completely
the inverse problem. We note that the elaborations done in equations
(\ref{19}) are, in fact, not necessary, since the torque can be immediately
inferred from $t_{ij}=f_{i}h_{j}$ by $f_{i}\simeq2\mu Su_{i}^{0}/l$
and $h_{j}\simeq ls_{j}$; we get $t_{ij}\simeq2\mu Va_{i}s_{j}$. 

We note here the representation
\begin{equation}
u_{ij}^{0}=\frac{1}{2}\left(s_{i}a_{j}+a_{i}s_{j}\right)=\frac{1}{4\mu V}M_{ij}\label{28}
\end{equation}
 for the focal strain, which follows immediately from the considerations
made above on the geometry of the focal region. This equation relates
the focal strain to the seismic moment; it may be used for assessing
the accumulation rate of the seismic moment from measurements of the
surface strain rate.\cite{key-23,key-24} 

It is worth noting that the estimations made above are affected by
an order-of-magnitude error in the numerical factors; this error is
related to the parameters $T$, $l$, the estimation of the derivatives
$\partial_{j}\delta$, the assumption $l_{1}=l_{2}=d=l$, the volume
$V=l^{3},$ etc. These errors affect mainly the volume $V$ in equations
(\ref{26}) and (\ref{27}). The errors in the seismic-moment parameters,
especially those related to noise, have been analzyed recently in
Ref. \cite{key-25}.

\section{Solution of the inverse problem}

Making use of the reduced moment $m_{ij}=M_{ij}/2\mu V$ and $m_{i}=M_{i}/2\mu V=M_{ij}n_{j}/2\mu V$,
equation (\ref{27}) leads to
\begin{equation}
s_{i}(\mathbf{na})+a_{i}(\mathbf{ns})=m_{i}\:\:;\label{29}
\end{equation}
using equations (\ref{10}) and (\ref{25}) the components $m_{i}$
of the reduced moment are given by
\begin{equation}
m_{i}=-\frac{T^{2}}{2R}\cdot\frac{c_{l}^{3}v_{li}+c_{t}^{3}v_{ti}}{c_{l}v_{l}^{2}+c_{t}v_{t}^{2}}\:\:.\label{30}
\end{equation}

We solve here the equations (\ref{29}) for the unit vectors $\mathbf{a}$
and $\mathbf{s}$, subject to the conditions
\begin{equation}
s_{i}^{2}=a_{i}^{2}=1\:\:,\:\: s_{i}a_{i}=0\:\:.\label{31}
\end{equation}
Since $M_{0}=0$ and $M^{2}>M_{4}^{2}$, we have $m_{0}=m_{ii}=0$
and $m^{2}>m_{4}^{2}$ (where $m_{4}=m_{ij}n_{i}n_{j}$ and $m^{2}=m_{i}^{2}$).
From equation (\ref{30}) we have $m_{i}<0$. The compatibility condition
$m^{2}>m_{4}^{2}$ can be checked immediately from equation (\ref{30})
(it arises from $v_{t}^{2}>0$). We write equations (\ref{29}) as
\begin{equation}
\alpha\mathbf{s}+\beta\mathbf{a}=\mathbf{m}\:\:\:,\label{32}
\end{equation}
where we introduce two new notations $\alpha=(\mathbf{na})$ and $\beta=(\mathbf{ns})$.
We assume that the vectors $\mathbf{s}$, $\mathbf{a}$ and $\mathbf{n}$
lie in the same plane, \emph{i.e. 
\begin{equation}
\beta\mathbf{s}+\alpha\mathbf{a}=\mathbf{n}\:\:.\label{33}
\end{equation}
}This condition determines the system of equations and ensures the
covariance of the solution; it is the covariance condition. From equations
(\ref{32}) and (\ref{33}) we get
\begin{equation}
2\alpha\beta=m_{4}\:\:,\:\:\alpha^{2}+\beta^{2}=m^{2}=1\:\:.\label{34}
\end{equation}

The equality $m^{2}=1$ (covariance condition) has important consequences;
it implies $M^{2}=(2\mu V)^{2}$, such that we can write the seismic
moment from equation (\ref{27}) as
\begin{equation}
M_{ij}=M\left(s_{i}a_{j}+a_{i}s_{j}\right)\:\:;\label{35}
\end{equation}
it follows the magnitude of the seismic moment $\left(M_{ij}{}^{2}\right)^{1/2}=\sqrt{2}M$;\cite{key-26}
$M$ is the magnitude of the \textquotedbl{}projection\textquotedbl{}
of the seismic-moment tensor along the observation radius. In addition,
from $E=W=\mu V$ (equation (\ref{24})) we have $E=M/2=\left(M_{ij}{}^{2}\right)^{1/2}/2\sqrt{2}$.
The magnitude $\left(M_{ij}{}^{2}\right)^{1/2}=\sqrt{2}M=2\sqrt{2}E$
may be used in the Gutenberg-Richter relation $\lg\left(M_{ij}{}^{2}\right)^{1/2}=1.5M_{w}+16.1$,
which defines the magnitude $M_{w}$ of the earthquake; in terms of
the earthquake energy this relation becomes $\lg E=1.5(M_{w}-\lg2)+16.1$
(where $\lg2\simeq0.3$). We note that an error of an order of magnitude
in the seismic moment ($M$, $E$, $\left(M_{ij}{}^{2}\right)^{1/2}$)
induces an error $\simeq0.3$ in the magnitude $M_{w}$. 

Further, from equation (\ref{30}), the equality $m^{2}=1$ can be
written as
\begin{equation}
\frac{T^{4}}{4R^{2}}\cdot\frac{c_{l}^{6}v_{l}^{2}+c_{t}^{6}v_{t}^{2}}{\left(c_{l}v_{l}^{2}+c_{t}v_{t}^{2}\right)^{2}}=1\:\:\:,\label{36}
\end{equation}
which gives the duration $T$ in terms of the displacements $v_{l,t}$
measured at distance $R$. Inserting $T$ in equation (\ref{26}),
we get
\begin{equation}
V^{2}=\frac{8\pi^{2}R^{3}}{c_{t}^{4}}\left(c_{l}v_{l}^{2}+c_{t}v_{t}^{2}\right)\left(c_{l}^{6}v_{l}^{2}+c_{t}^{6}v_{t}^{2}\right)^{1/2}\label{37}
\end{equation}
and the magnitude of the seismic moment and the energy of the earthquake
\begin{equation}
M=2E=2\mu V=2\pi\rho(2R)^{3/2}\left(c_{l}v_{l}^{2}+c_{t}v_{t}^{2}\right)^{1/2}\left(c_{l}^{6}v_{l}^{2}+c_{t}^{6}v_{t}^{2}\right)^{1/4}\label{38}
\end{equation}
 in terms of the displacements $v_{l,t}$ measured at distance $R$.
In addition, eliminating $R^{2}$ between equations (\ref{26}) and
(\ref{36}) we can express the focal volume as
\begin{equation}
V=\frac{\pi T^{3}}{c_{t}^{2}}\cdot\frac{c_{l}^{6}v_{l}^{2}+c_{t}^{6}v_{t}^{2}}{c_{l}v_{l}^{2}+c_{t}v_{t}^{2}}\:\:.\label{39}
\end{equation}
 The solutions of the system of equations (\ref{34}) are given by
\begin{equation}
\alpha=\sqrt{\frac{1+\sqrt{1-m_{4}^{2}}}{2}}\:\:,\:\:\beta=sgn(m_{4})\sqrt{\frac{1-\sqrt{1-m_{4}^{2}}}{2}}\label{40}
\end{equation}
 and $\alpha\longleftrightarrow\pm\beta$, $\alpha,\,\beta\longleftrightarrow-\alpha,\,-\beta$.
Making use of equations (\ref{30}) and (\ref{36}), the parameters
$m_{i}$ and $m_{4}$ are given by
\begin{equation}
m_{i}=-\frac{c_{l}^{3}v_{li}+c_{t}^{3}v_{ti}}{\left(c_{l}^{6}v_{l}^{2}+c_{t}^{6}v_{t}^{2}\right)^{1/2}}\:\:,\:\: m_{4}=-\frac{c_{l}^{3}(\mathbf{v}_{l}\mathbf{n})}{\left(c_{l}^{6}v_{l}^{2}+c_{t}^{6}v_{t}^{2}\right)^{1/2}}\:\:.\label{41}
\end{equation}
 Finally, we get the vectors
\begin{equation}
\begin{array}{c}
\mathbf{s}=\frac{\alpha}{\alpha^{2}-\beta^{2}}\mathbf{m}-\frac{\beta}{\alpha^{2}-\beta^{2}}\mathbf{n}\:\:,\\
\\
\mathbf{a}=-\frac{\beta}{\alpha^{2}-\beta^{2}}\mathbf{m}+\frac{\alpha}{\alpha^{2}-\beta^{2}}\mathbf{n}\:\:;
\end{array}\label{42}
\end{equation}
from equations (\ref{32}) and (\ref{33}); these solutions are symmetric
under the operations $\mathbf{s}\longleftrightarrow\mathbf{a}$ ($\alpha\longleftrightarrow-\beta$)
and $\mathbf{s}\longleftrightarrow-\mathbf{a}$ ($\alpha\longleftrightarrow\beta,$
or $\alpha,\,\beta\longleftrightarrow-\alpha,\,-\beta$). The seismic
moment given by equation (\ref{35}) is determined up to these symmetry
operations.
\begin{figure}
\noindent \begin{centering}
\includegraphics[clip,scale=0.8]{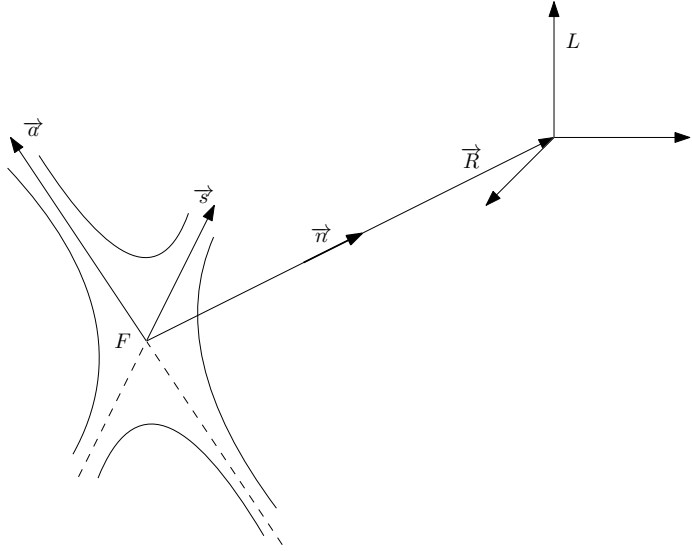}
\par\end{centering}

\caption{The hyperbola of the displacement ($\mathbf{a}$) in the fault plane
(fault direction $\mathbf{s}$) at the focus ($\mathbf{F}$), seen
from the local frame $L$. \label{fig: Fig.3}}

\end{figure}
 We can see that the seismic-moment tensor given by eqaution (\ref{35})
is determined by $M$ (equation (\ref{38})) and the vectors $\mathbf{s}$
and $\mathbf{a}$ given by equations (\ref{42}), with the coefficients
$\alpha,\,\beta$ given by equations (\ref{40}); the vector $\mathbf{n}$
is known and the vector $\mathbf{m}$ and the scalar $m_{4}$ are
given by the experimental data (equations (\ref{41})). Equations
(\ref{42}) are manifestly covariant.

The eigenvalues of the seismic moment given by equation (\ref{35})
are $\pm M$ (we leave aside the eigenvalue zero); the corresponding
eigenvectors $\mathbf{w}$ are given by $\mathbf{aw}=\pm\mathbf{sw}$,
which imply $\mathbf{mw}=\pm\mathbf{nw}$; the vectors $\mathbf{w}$
are directed along the bisectrices of the angles made by $\mathbf{s}$
and $\mathbf{a}$, or $\mathbf{m}$ and $\mathbf{n}$ ($\mathbf{w}\sim\mathbf{s}\pm\mathbf{a}$).
The associated quadratic form $M_{ij}x_{i}x_{j}=const$ is a rectangular
hyperbola in the reference frame defined by the vectors $\mathbf{s}$
and $\mathbf{a}$; by using the coordinates $u=\mathbf{sx}$ and $v=\mathbf{ax}$
in equation (\ref{35}), the equation of this hyperbola is $uv=const/2M.$
Actually, in the local frame (coordinates $x_{i}$), the quadratic
form $M_{ij}x_{i}x_{j}=const$ is a degenerate hyperboloid, consisting
of a family of parallel hyperbolas displaced along the third axis
(perpendicular to the $u$- and $v$-axes). Making use of equations
(\ref{35}) and (\ref{42}), this quadratic form can also be written
as
\begin{equation}
2\xi\eta-m_{4}\left(\xi^{2}+\eta^{2}\right)=const\:\:\:,\label{43}
\end{equation}
where the coordinates $\xi=m_{i}x_{i}$ and $\eta=n_{i}x_{i}$ are
directed along the vectors $\mathbf{m}$ and $\mathbf{n}$, respectively.
The asymptotics of this hyperbola are $\xi=m_{4}\eta/\left(1+\sqrt{1-m_{4}^{2}}\right)$
and $\eta=m_{4}\xi/\left(1+\sqrt{1-m_{4}^{2}}\right)$ (corresponding
to the asymptotics $u=(\alpha\xi-\beta\eta)/(\alpha^{2}-\beta^{2})=0$
and $v=(-\beta\xi+\alpha\eta)/(\alpha^{2}-\beta^{2})=0$). (Fig. \ref{fig: Fig.3})

Finally, by making use of equations (\ref{42}) in equation (\ref{35})
we get the solution for the seismic moment
\begin{equation}
M_{ij}=\frac{M}{1-m_{4}^{2}}\left[m_{i}n_{j}+m_{j}n_{i}-m_{4}\left(m_{i}m_{j}+n_{i}n_{j}\right)\right]\:\:\:,\label{44}
\end{equation}
where $M$ is given by equation (\ref{38}) and $m_{i},\, m_{4}$
are given by equations (\ref{41}); the focal strain is $u_{ij}^{0}=M_{ij}/2M$
(equation (\ref{28})). In equation (\ref{44}) there are only three
independent components of the seismic tensor, according to the equations
$m_{ij}n_{j}=m_{i}$ ($m_{ij}=M_{ij}/M$): the vectors $\mathbf{n}$
and $\mathbf{m}$ are known (equation (\ref{41})) from experimental
data, such that these equations can be viewed as three conditions
imposed upon the six components $M_{ij}.$ Also, we can see that there
exist only three independent components of the seismic tensor $M_{ij}$
from the conditions $M_{0}=M_{ii}=0$, $M_{ij}s_{j}s_{i}=0$ (or $M_{ij}a_{i}a_{j}=0$)
and $m_{i}^{2}=1$. The later equality arises from the covariance
condition, which, together with the energy conservation, determines
the duration $T$ of the earthquake, the volume $V$ of the focal
region and the magnitude parameter $M$ of the seismic moment.

\section{Isotropic seismic moment}

An isotropic seismic moment $M_{ij}=-M\delta_{ij}$ is an interesting
particular case, since it can be associated with seismic events caused
by explosions.\cite{key-27} In this case the transverse displacement
is vanishing ($\mathbf{u}_{t}^{n,f}=0$), $\mathbf{M}=-M\mathbf{n}$,
$M_{4}=-M$ and $\mathbf{v}_{l}=(R/c_{l}T)\mathbf{u}_{l}^{n}$ (equations
(\ref{7}) and (\ref{9})); from equations (\ref{10}) and (\ref{17})
we get
\begin{equation}
\mathbf{M}=-4\pi\rho TRc_{l}^{3}\mathbf{v}_{l}\:\:,\:\: E=\frac{4\pi\rho R^{2}}{T}c_{l}v_{l}^{2}\label{45}
\end{equation}
we can see that $\mathbf{v}_{l}\mathbf{n}>0$ corresponds to $M>0$
(explosion), while the case $\mathbf{v}_{l}\mathbf{n}<0$ corresponds
to an implosion. The focal zone is a sphere with radius of the order
$l$, and the vectors $\mathbf{s}$ and $\mathbf{a}$ are equal ($\mathbf{s}=\mathbf{a}$)
and depend on the point on the focal surface; the magnitude of the
focal displacement is $u^{0}=l$. The considerations made above for
the geometry of the focal region lead to the representation
\begin{equation}
M_{ij}=-2V(2\mu+\lambda)\delta_{ij}=-2\rho c_{l}^{2}V\delta_{ij}\:\:\:,\label{46}
\end{equation}
where $V=Sl$ denotes the focal volume and $S$ is the area of the
focal region (we note that $t_{ij}$ changes sign in equation (\ref{20})).
Similarly, the energy is $E=W=\frac{1}{2}M$ ($M>0$), such that,
making use of equations (\ref{45}), we get $c_{l}T=\sqrt{2Rv_{l}}$,
\begin{equation}
M=2\pi\rho c_{l}^{2}(2Rv_{l})^{3/2}=2\rho c_{l}^{2}V\:\:,\label{47}
\end{equation}
 and the focal volume $V=\pi(2Rv_{l})^{3/2}$. These equations determine
the seismic moment and the volume of the focal region from the displacement
$v_{l}$ measured at distance $R$. A superposition of shear faulting
and isotropic focal mechanisms cannot be resolved, because the longitudinal
displacement $\mathbf{v}_{l}$ includes indiscriminately contributions
from both mechanisms.

\section{Discussion and concluding remarks}

We can summarize the results as follows. Making use of the longitudinal
displacement $\mathbf{v}_{l}$ and the transverse displacement $\mathbf{v}_{t}$,
measured at the Earth's surface, we compute the magnitude parameter
$M$ from equation (\ref{38}) and the vector $\mathbf{m}$ and the
scalar $m_{4}$ from equation (\ref{41}); then, from equation (\ref{44})
we get the seismic moment $M_{ij}$. The energy released by the earthquake
is $E=M/2$ and an estimate of the focal volume is given by $V=M/2\rho c_{t}^{2}$
(equations (\ref{27}) and (\ref{35})). An estimation of the duration
$T$ of the earthquake is provided by equation (\ref{36}). The focal
slip is of the order $V^{1/3}$ and the focal strain is of the order
$M_{ij}/2M$ (equation (\ref{28})). From the magnitude $\left(M_{ij}{}^{2}\right)^{1/2}=\sqrt{2}M$
of the seismic moment we may estimate the magnitude $M_{w}$ of the
earthquake by means of the Gutenberg-Richter relation. A similar procedure
holds for an isotropic seismic moment (preceding section). 

Making use of $\mathbf{m}$ and $m_{4}$ in equations (\ref{42})
we compute the normal $\mathbf{s}$ to the fault plane and the unit
slip vector $\mathbf{a}$ in the fault plane; the quadratic form associated
to the seismic moment is a degenerate hyperboloid which reduces to
a hyperbola in the $(\mathbf{s},\mathbf{a})$-plane with asymptotics
along the vectors $\mathbf{s}$ and $\mathbf{a}$. This hyperbola
is tighter (closer to the origin) for higher $M$. 

It is convenient to have an estimation of the order of magnitude of
the various quantities introduced in this paper. To this end we use
a generic velocity $c$ for the seismic waves and a generic vector
$\mathbf{v}$ for the displacement in the far-field seismic waves.
Equation (\ref{36}) (which is $m^{2}=1$) gives $cT\simeq\sqrt{2Rv}$,
which provides an estimate of the duration of the earthquake in terms
of the displacement measured at distance $R$. The focal volume can
be estimated from equation (\ref{26}) as $V\simeq\pi\left(2Rv\right)^{3/2}\simeq\pi(cT)^{3}$,
as expected (dimension $l$ of the focal region of the order $cT$;
the rate of the focal slip is $l/T\simeq c$). Also, from equation
(\ref{38}) we have the energy $E\simeq\mu V\simeq M/2\simeq2\rho c^{2}V$,
where $M$ is related to the magnitude $\left(M_{ij}^{2}\right)^{1/2}=\sqrt{2}M$
of the seismic moment (and the magnitude of the vector $M_{ij}n_{j}$).
From equation (\ref{28}) we get a focal strain of the order unity,
as expected. 

In conclusion, it is shown in this paper that the displacement in
the far-field seismic waves provides information about the structure
of the focal region; in particular, this displacement can be employed
to determine the seismic-moment tensor for a fault slip, localized
both in space and time (the inverse problem in Seismology). In this
case the vectorial (Kostrov) representation of the seismic moment
(dyadic representation) is written with four (unknown) parameters;
one is the magnitude of the focal displacement, while the other three
define the spatial orientation of the seismic tensor (orientation
of the fault and the displacement direction). These unknown parameters
are determined from the three equations relating the far-field displacement
to the seismic tensor and the equation which relates the energy released
in the earthquake (and carried by the seismic waves) to the focal
displacement (and the fault focal volume), via the mechanical work
done in the focal region, together with the covariance condition.
The solution of the resulting system of equations makes the graphical
representation of the quadratic form associated to the seismic-moment
tensor, which is a hyperbola, to offer a (three-dimensional) image
of the focal region. The asymptotics of the hyperbola give the direction
of the focal displacement and the orientation of the fault (seismic
hyperbola). Besides solving the inverse problem in Seismology for
a localized fault slip, the geometry of the fault focal region (which
leads to Kostrov representation) and the displacement in the far-field
seismic waves provide reasonable estimations of the fault focal volume,
focal strain, duration and energy of the earthquake and magnitude
of the seismic moment. Also, the special case of an isotropic seismic
moment is presented. More complex situations, like a superposition
of point-like faults, or a combination of point-like faults and isotropic
and dipole components imply more than four unknowns in the seismic
tensor; since we have only four equations, the inverse problem in
such cases is undetermined, within the present procedure. The procedure
presented in this work makes use of manifestly covariant expressions
of the data for determining the seismic moment.

Finally, we note that a similar deduction of the seismic-moment tensor
can be done by using the (quasi)-static displacement at Earth's surface,
derived in Ref. \cite{key-28}; since it implies a specific treatment,
its presentation is deferred to a forthcoming publication. 

\textbf{Acknowledgments.} The author is indebted to his colleagues
in the Department of Engineering Seismology, Institute of Earth\textquoteright{}s
Physics, Magurele-Bucharest, for many enlightening discussions, and
to the members of the Laboratory of Theoretical Physics at Magurele-Bucharest
for many useful discussions and a throughout checking of this work.
This work was partially supported by the Romanian Government Research
Grant \#PN16-35-01-07/11.03.2016.

\end{document}